\documentclass[preprint2]{aastex}

\usepackage{lscape}
\usepackage{graphicx}
\usepackage{txfonts}
\usepackage{natbib}
\usepackage[colorlinks, citecolor={blue}]{hyperref}
\def\arxivprefixe{arXiv}
\def\arxivprefixesep{:}

\newcommand{\eprint}[2][]{%
{\tt\if!#1!#2\else#1\arxivprefixesep\ignorespaces#2\fi}%
}

\newcommand{\myemail}{maria.argudo@uantof.cl}

\shorttitle{LSSGALPY}
\shortauthors{Argudo-Fern\'andez et al.}

\begin{document}


\title{LSSGalPy\thanks{Code and tables at \href{https://github.com/margudo/LSSGALPY}{https://github.com/margudo/LSSGALPY}} : Interactive Visualization of the Large-scale Environment Around Galaxies}


\author{M. Argudo-Fern\'andez \altaffilmark{1}, S. Duarte Puertas\altaffilmark{2}, J. E. Ruiz\altaffilmark{2}, J. Sabater\altaffilmark{3}, S. Verley\altaffilmark{4,5}, and G. Bergond\altaffilmark{6}}

\altaffiltext{1}{Universidad de Antofagasta, Unidad de Astronom\'ia, Facultad Cs. B\'asicas, Av. U. de Antofagasta 02800, Antofagasta, Chile, \myemail}
\altaffiltext{2}{Instituto de Astrof\'isica de Andaluc\'ia (CSIC) Apdo. 3004, 18080 Granada, Spain}
\altaffiltext{3}{Institute for Astronomy, University of Edinburgh, Edinburgh EH9 3HJ, UK}
\altaffiltext{4}{Departamento de F\'isica Te\'orica y del Cosmos Universidad de Granada, 18071 Granada, Spain}
\altaffiltext{5}{Instituto Universitario Carlos I de Física Teórica y Computacional, Universidad de Granada, 18071 Granada, Spain}
\altaffiltext{6}{Centro Astron\'omico Hispano Alemán, Compl. Observatorio Calar Alto s/n, Sierra de los Filabres, 04550 Gerg\'al, Spain}


\begin{abstract}
New tools are needed to handle the growth of data in astrophysics delivered by recent and upcoming surveys. We aim to build open-source, light, flexible, and interactive software designed to visualize extensive three-dimensional (3D) tabular data. Entirely written in the Python language, we have developed interactive tools to browse and visualize the positions of galaxies in the universe and their positions with respect to its large-scale structures (LSS). Motivated by a previous study, we created two codes using Mollweide projection and wedge diagram visualizations, where survey galaxies can be overplotted on the LSS of the universe. These are interactive representations where the visualizations can be controlled by widgets. We have released these open-source codes that have been designed to be easily re-used and customized by the scientific community to fulfill their needs. The codes are adaptable to other kinds of 3D tabular data and are robust enough to handle several millions of objects.
\end{abstract}





\section{Introduction}

The amount of available astronomical data is growing exponentially due to the increasing number of observing facilities, automated surveys, and simulations. The complexity of these data is also increasing with time. For instance, where the positions of stars were only available as two spatial coordinates before, recent automated surveys such a Gaia \citep{2016A&A...595A...1G} provide not only the spatial coordinates, but also the parallaxes and proper motions of the stars along with a bundle of photometric data. The tools to handle astronomical data have to ensue with this technical evolution and be able to deal with several tens of millions objects and represent them in a flexible, interactive way.

During the last decade, tools for data inspection and visualization have been developed to follow the evolution of the astrophysical data \citep{2008NJPh...10l5015S,2011PASA...28..150H,2012AN....333..505G}. Initiatives like the International Virtual Observatory Alliance (\href{www.ivoa.net}{\texttt{www.ivoa.net}}) have arisen to homogenize astronomical data representation and exchange, with tools like Aladin \citep{2000A&AS..143...33B}, or the Visualization Interface for the Virtual Observatory \citep[VisIVO;][]{2007PASP..119..898C,2015A&C....11..146S}. In particular, very dedicated and flexible tools have greatly simplified everyday tasks for data manipulation, which is often quite cumbersome, such as cross-matching catalogs or rapid subsample filtering, browsing, and visualization of multidimensional tabular data sets \citep[see \textsc{topcat} for instance,][]{2005ASPC..347...29T}. Three-dimensional (3D) visualizations have also experienced great advances, benefiting from the inclusion of mature tools such as \textsc{Blender} \citep{2013PASP..125..731K,2016A&C....15...50N}.

However, simple and flexible codes can also be valid for specific tasks. The multiplication of individual codes can be either overwhelming or not being noticed by the researcher, lost among the large number of those available; they also may be of invaluable help to get started on handling a specific task. To be useful, these codes have to be available as free software and written in a programming language that is well spread throughout the community so that they can be easily re-used and modified according to specific needs \citep{2014PLSCB..10E3542G}. In this regard, Python has gained recognition during the last 15 years in the astrophysical community because it is a simple, easy to learn, and powerful language. It is particularly well suited for research due to its large collection of libraries and flexibility. The popularity of \textsc{Astropy} \citep{2013A&A...558A..33A} is rising rapidly and its affiliated packages are a demonstration of it.

The objectives of the paper are to present \textsc{LSSGalPy}, an interactive visualization of the large-scale environment around galaxies, to encourage its use, and to foster re-use and modifications by other authors. First, we present the software in Section~\ref{sec:software} and mention how and where to obtain the code. Emphasis has been made to create a very short, simple, robust, and easy to modify code to encourage its adaptation and re-use in the community. Next, in Section~\ref{sec:uses}, we present the particular case for which \textsc{LSSGalPy} has been built. We also mention some other cases where this code has been used and offer examples of some other contexts that may benefit from \textsc{LSSGalPy} after minor modifications to the code. We conclude with a brief summary in Section~\ref{sec:summary}.

\section{Software Presentation} \label{sec:software}

\textsc{LSSGalPy} provides visualization tools to compare the 3D positions (right ascension, declination, and redshift) of one or more samples of isolated systems with respect to the locations of the large-scale structures (LSS) of the universe (see Figs.~\ref{fig:mollweide} and \ref{fig:wedge}). The basic functionality of these interactive tools is the use of different projections to study the relation of the galaxies with their local environment and with the LSS. Additionally, we may seamlessly add or remove samples, change the marker size and/or symbol, modify transparency, etc. These tools have been tested using up to 30 million galaxies and still work perfectly and very smoothly on any standard laptop.

During the writing of the software, emphasis was made on obtaining a high-quality code. The code is compact (less than 100 lines of code) and has an easy to read structure through list comprehension and extensive use of already existing functions in the imported Python libraries. The code is designed to allow easy modifications, which is useful to the scientific community since users can use their own galaxy catalogs. We have developed the code under a Linux platform, but it may be straightforwardly ported to any other operating system.

The software is primarily available for download at \href{https://github.com/margudo/LSSGALPY}{GitHub}\footnote{\href{https://github.com/margudo/LSSGALPY}{\texttt{https://github.com/margudo/LSSGALPY}}}. To install the software on your local computer, simply copy all files\footnote{The input files are in ascii format. The total size is smaller than 20\,MB.} in a directory and run the Python files \textit{LSSGALPY\_mollweide.py} or \textit{LSSGALPY\_wedge.py} to visualize the LSS environment using a Mollweide projection or a wedge diagram, respectively. The Mollweide projection shows the location of the galaxies in the sky with respect to their right ascension and declination, and complementary, a wedge diagram shows the location with respect to right ascension and redshift. The Python2 and Jupyter notebook versions are available in the GitHub repository. These Python2 tools currently work as widgets in a local computer but the Jupyter notebook versions may be also accessed through a browser and executed in the \href{http://mybinder.org/repo/margudo/lssgalpy}{myBinder}\footnote{\href{http://mybinder.org/repo/margudo/lssgalpy}{\texttt{http://mybinder.org/repo/margudo/lssgalpy}}} cloud infrastructure. The myBinder repository allows the interactive inspection of the richly documented code as well as testing its functionalities without altering the local execution environment.

The required dependencies are
\begin{itemize}
\item \href{http://www.numpy.org}{\textsc{NumPy}\footnote{\href{http://www.numpy.org}{\texttt{http://www.numpy.org}}}:} array processing for numbers, strings, records, and objects;
\item \href{http://matplotlib.org}{\textsc{matplotlib}\footnote{\href{http://matplotlib.org}{\texttt{http://matplotlib.org}}}:} Python 2D plotting library; and
\item \href{http://matplotlib.org/basemap}{basemap\footnote{\href{http://matplotlib.org/basemap}{\texttt{http://matplotlib.org/basemap}}}:} add-on toolkit for matplotlib.
\end{itemize}

The code is licensed under \href{https://opensource.org/licenses/MIT}{MIT License (MIT\footnote{\href{https://opensource.org/licenses/MIT}{\texttt{https://opensource.org/licenses/MIT}}})}. The code is also shared via repositories such as \href{http://ascl.net/1505.012}{Astrophysics Source Code Library (ASCL\footnote{\href{http://ascl.net/1505.012}{\texttt{http://ascl.net/1505.012}}})} and \href{https://zenodo.org/record/17512\#.V9pOXHQyrMU}{Zenodo\footnote{\href{https://zenodo.org/record/17512\#.V9pOXHQyrM}{\texttt{https://zenodo.org/record/17512\#.V9pOXHQyrMU}}}}.

Some tutorials showcasing the interactive features of the software are available as video demonstrations on Vimeo:
\begin{itemize}
\item \href{https://vimeo.com/133013373}{\textsc{LSSGalPy}: Mollweide projection\footnote{\href{https://vimeo.com/133013373}{\texttt{https://vimeo.com/133013373}}}} shows how to explore the location of different samples of galaxies (samples explained in Sect.~\ref{sec:uses}) with respect to third dimension (redshift) in the Mollweide projection (right ascension and declination) visualization; and
\item \href{https://vimeo.com/133013372}{\textsc{LSSGalPy}: Wedge diagram\footnote{\href{https://vimeo.com/133013372}{\texttt{https://vimeo.com/133013372}}}} similarly shows how to explore the location of galaxies with respect to the declination in the wedge diagram (right ascension and redshift) visualization. This demonstration also shows how to drag and zoom in a particular region, which corresponds to the Coma cluster.
\end{itemize}

\section{Use Cases} \label{sec:uses}

\subsection{Principal Use Case}

\begin{landscape}
\begin{figure}
\centering
\includegraphics[width=1.35\textwidth]{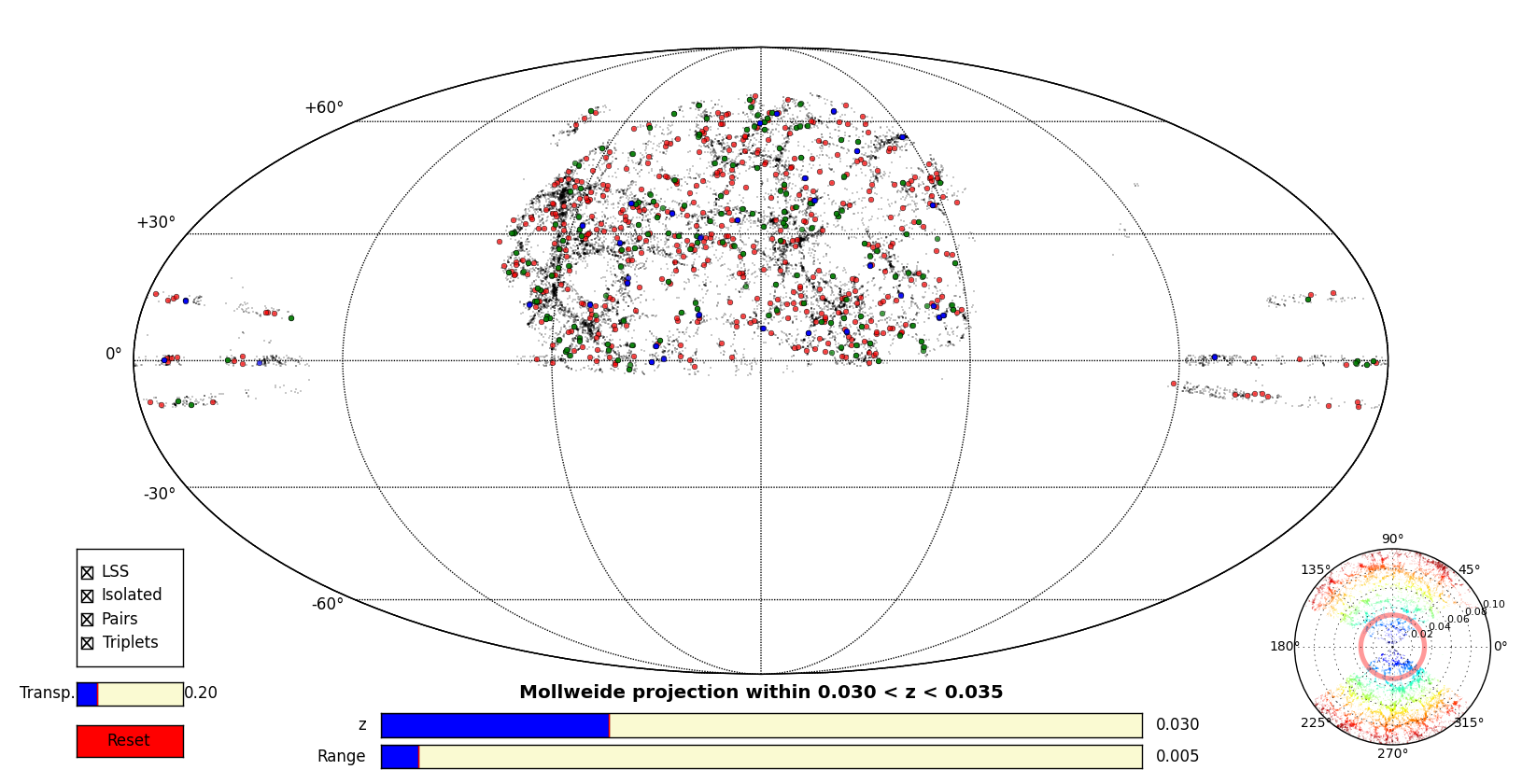}
\caption{Interactive 3D visualization software: Mollweide projection. Mollweide projection of galaxies in the LSS (black points) in the redshift range $0.030 < z < 0.035$ (default values according to the blue bars in the lower part of the figure). Isolated galaxies within the same redshift range are depicted by red circles. Green and blue circles represent isolated pairs and isolated triplets, respectively. The samples can be toggled on and off by selecting in the samples box located in the left part of the figure. The transparency of the represented redshift range can be modified by sweeping the blue bar under the samples box. To guide the eye, we show a wedge diagram for LSS galaxies within -2 and 2 degrees in declination in the right lower part of the figure. The red ring shows the selected redshift range in the central Mollweide projection. Color code of the complementary diagram corresponds to the redshift from $z=0$ (blue) to $z=0.10$ (red). The reset button returns to the default values.A color version of this figure is available in the online journal.}
\label{fig:mollweide}
\end{figure}
\end{landscape}

\begin{landscape}
\begin{figure}
\centering
\includegraphics[width=1.35\textwidth]{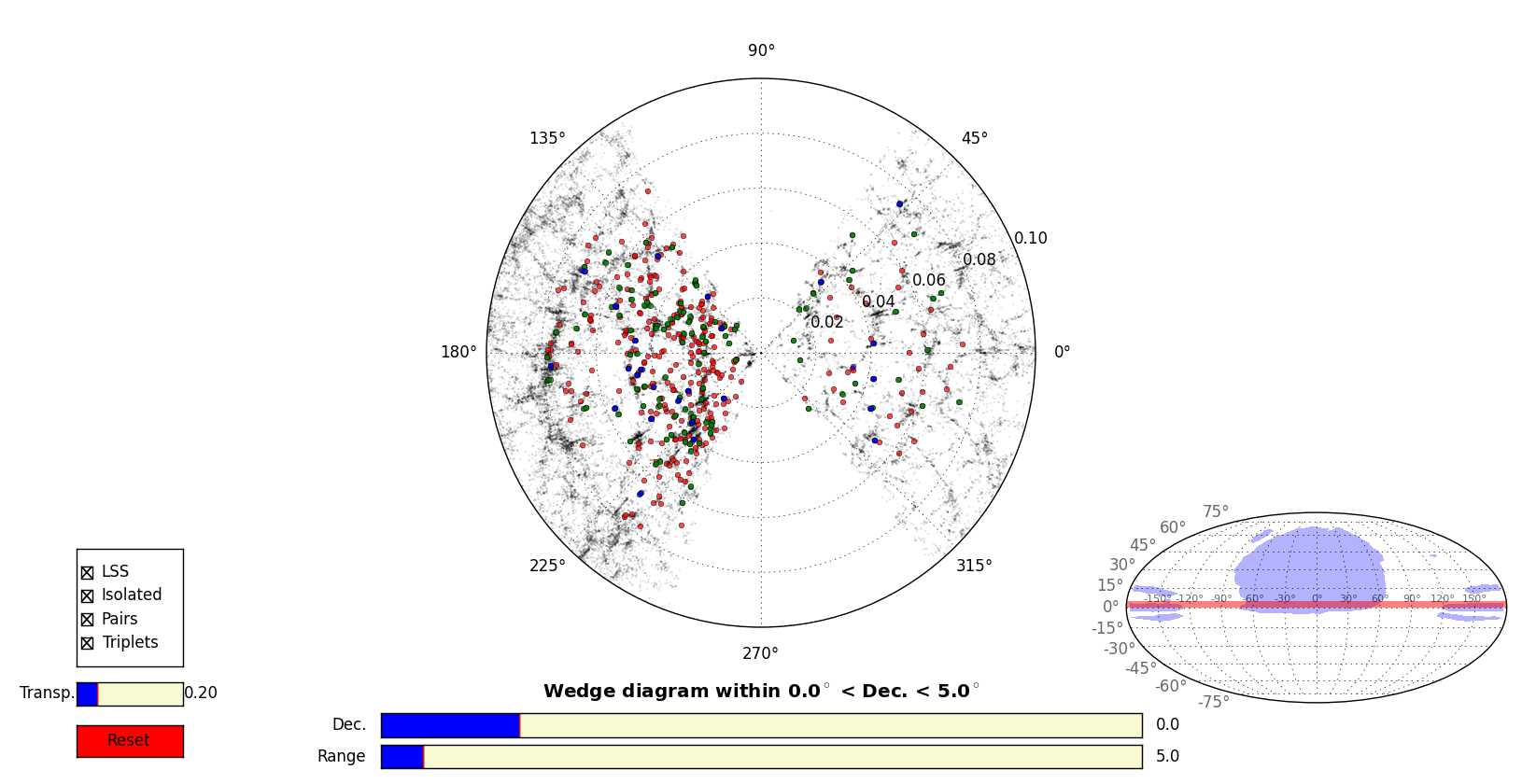}
\caption{Interactive 3D visualization software: wedge diagram. Wedge diagram of galaxies in the LSS (black points) within 0 and 5 degrees in declination (default values according to the blue bars in the lower part of the figure). The declination range can be changed using the blue bars in the lower part of the screen. As in Fig.~\ref{fig:mollweide}, isolated galaxies are depicted by red circles, and green and blue circles represent isolated pairs and isolated triplets, respectively. The samples can be toggled on and off by selecting in the samples box. The transparency of the represented redshift range can be modified by sweeping the blue bar under the samples box. To guide the eye, we show the Mollweide projection for LSS galaxies in the SDSS footprint, where the red stripe shows the selected declination range in the central wedge diagram.A color version of this figure is available in the online journal.}
\label{fig:wedge}
\end{figure}
\end{landscape}

The codes were primarily designed and developed for the article by \citet{2015A&A...578A.110A}. We visualized how the isolated galaxies (hereafter SIG for {\bf S}DSS-based catalog of {\bf I}solated {\bf G}alaxies), isolated pairs (hereafter SIP for {\bf S}DSS-based catalog of {\bf I}solated {\bf P}airs), and isolated triplets are related to the galaxies in their LSS. In particular, we used a Mollweide projection in combination with a wedge diagram (see Fig.~\ref{fig:mollweide}) and vice versa (Fig.~\ref{fig:wedge}). Note that the code could also work with several tens of other types of projections. We used the blue bars displayed under the sky map in Fig.~\ref{fig:mollweide} to visualize the locations of the galaxies in our study for different values of redshifts and redshift ranges. Similarly, using the bars below wedge diagram (in Fig.~\ref{fig:wedge}), we explored the locations of the galaxies for different values of the declinations and declination ranges. 

As a result, we observed that most of the isolated galaxies, isolated pairs, and isolated triplets are related more to the outer parts of filaments, walls, and clusters, and generally differ of the void population of galaxies. In fact, only one-third of SIG galaxies are located in voids. Also, using \textsc{LSSGalPy}, we checked that galaxies with low values of the tidal strength parameter, $Q_{\rm LSS}$ \citep{2007A&A...472..121V,2015A&A...578A.110A}, are mainly located in void regions, and galaxies with higher $Q_{\rm LSS}$ are more closely related to denser structures, such as the filaments or walls defining the LSS of the universe.

\subsection{Other Uses}

\citet{2016A&A...592A..30A} investigated the effect of local and large-scale environments on nuclear activity and star formation. For the purpose of this study, we compared the positions of active galactic nuclei (AGN) and star forming galaxies in the SIG and SIP samples according to three different ranges of values of their $Q_{\rm LSS}$. We observed that the fraction of optical AGN for high-mass SIG and SIP galaxies increases with denser large-scale environment, and on the contrary, the fraction of optical AGN for low-mass SIG and SIP galaxies decreases from voids to denser regions \citep{2016A&A...592A..30A}.

\textsc{LSSGalPy} has also been used in \citet{2016MNRAS.463..913J} and \citet{2016NatCo...713269C} to visualize the localization of galaxies with kinematically decoupled stellar and gaseous components, in conjunction with the estimation of the tidal strength parameter, $Q_{\rm LSS}$, affecting each galaxy. They found that there is a trend showing that the kinematically misaligned galaxies are more isolated than the aligned control sample. In fact most of the misaligned galaxies are mainly located in void regions, and control galaxies are more related with denser structures, such as the filaments or walls defining the LSS of the Universe.

\subsection{Suggestions for Other Possible Uses}

Although \textsc{LSSGalPy} has been developed for the needs of one particular initial work, care has been taken in its design to allow easy modifications in order to be able to visualize other kind of data. Here we suggest some of them.

The software can be straightforwardly adapted to the visualization of stellar surveys. The most ambitious is the Gaia mission \citep{2016A&A...595A...1G}, with its first data release, consisting of astrometry and photometry for over 1 billion sources and positions, parallaxes, and mean proper motions for a subset of two million of bright stars. By using the parallax instead of the redshift information, it will be possible to visualize the 3D positions of the stars in the Milky Way.

New releases of large 3D surveys may also benefit from the visualization capabilities of \textsc{LSSGalPy}, in particular the SDSS-DR13, first spectroscopic data from the SDSS-IV \citep{2016arXiv160802013S} with its dedicated surveys APOGEE-2, MaNGA, and eBOSS.

\textsc{LSSGalPy} may also be adapted to visualize data cubes, with some small modifications to explore to integral field spectroscopy data (using the velocity instead of third spatial axis). For instance, in the MaNGA integral field unit (IFU) spectra \citep{2015ApJ...798....7B, 2016AJ....152..197Y} where data are available for 1390 spatially resolved integral field unit observations of nearby galaxies in the SDSS-DR13. Other surveys such as CALIFA, the Calar Alto Legacy Integral Field Area survey \citep{2012A&A...538A...8S}, or other IFU surveys can also benefit from the \textsc{LSSGalPy} visualization possibilities.

More generally, \textsc{LSSGalPy} can be adapted to any data in three dimensions; the parameter space is not only restricted to spatial dimensions: time or any other quantity can as well be involved in those 3D data exploration widgets. \textsc{LSSGalPy} is also easily modifiable by using any of the \href{http://matplotlib.org/basemap/users/mapsetup.html}{24 different map projections available\footnote{\href{http://matplotlib.org/basemap/users/mapsetup.html}{\texttt{http://matplotlib.org/basemap/users/mapsetup.html}}}} of the projection and mapping \href{http://matplotlib.org/basemap/}{Matplotlib Basemap Toolkit\footnote{\href{http://matplotlib.org/basemap/}{\texttt{http://matplotlib.org/basemap/}}}}.

\section{Summary} \label{sec:summary}

In this paper we present \textsc{LSSGalPy}, an interactive visualization of the large-scale environment around galaxies. We provide two basic visualization codes, a Mollweide projection and a wedge diagram where some survey galaxies can be overplotted on the large-scale structures of the Universe. These are interactive representations and the visualizations can be controlled by widgets. The codes were designed to be easily modifiable and adaptable to other kinds of 3D data and re-used to suit the needs of the scientific community.

\begin{acknowledgements}
The authors acknowledge the anonymous referee for the very detailed and useful report, which helped to clarify and improve the quality of this work. M.A.F. is grateful for financial support from CONICYT FONDECYT project no. 3160304. This research made use of \textsc{Astropy}, a community-developed core \textsc{Python} ({\tt http://www.python.org}) package for Astronomy \citep{2013A&A...558A..33A}; \textsc{ipython} \citep{PER-GRA:2007}; \textsc{matplotlib} \citep{Hunter:2007}; \textsc{NumPy} \citep{:/content/aip/journal/cise/13/2/10.1109/MCSE.2011.37}; \textsc{scipy} \citep{citescipy}; \textsc{astroML} \citep{astroML}; and \textsc{topcat} \citep{2005ASPC..347...29T}.\\

Caveats: The authors are not responsible of potential content decay in complementary repository services (GitHub, Vimeo, Zenodo) or malfunctioning of the myBinder cloud infrastructure service.
\end{acknowledgements}

\bibliography{references}

\begin{thebibliography}{26}
\expandafter\ifx\csname natexlab\endcsname\relax\def\natexlab#1{#1}\fi

\bibitem[{{Argudo-Fern{\'a}ndez} {et~al.}(2016){Argudo-Fern{\'a}ndez}, {Shen},
  {Sabater}, {Duarte Puertas}, {Verley}, \& {Yang}}]{2016A&A...592A..30A}
{Argudo-Fern{\'a}ndez}, M., {Shen}, S., {Sabater}, J., {et~al.} 2016, \aap,
  592, A30

\bibitem[{{Argudo-Fern{\'a}ndez} {et~al.}(2015){Argudo-Fern{\'a}ndez},
  {Verley}, {Bergond}, {Duarte Puertas}, {Ramos Carmona}, {Sabater},
  {Fern{\'a}ndez Lorenzo}, {Espada}, {Sulentic}, {Ruiz}, \&
  {Leon}}]{2015A&A...578A.110A}
{Argudo-Fern{\'a}ndez}, M., {Verley}, S., {Bergond}, G., {et~al.} 2015, \aap,
  578, A110

\bibitem[{{Astropy Collaboration} {et~al.}(2013){Astropy Collaboration},
  {Robitaille}, {Tollerud}, {Greenfield}, {Droettboom}, {Bray}, {Aldcroft},
  {Davis}, {Ginsburg}, {Price-Whelan}, {Kerzendorf}, {Conley}, {Crighton},
  {Barbary}, {Muna}, {Ferguson}, {Grollier}, {Parikh}, {Nair}, {Unther},
  {Deil}, {Woillez}, {Conseil}, {Kramer}, {Turner}, {Singer}, {Fox}, {Weaver},
  {Zabalza}, {Edwards}, {Azalee Bostroem}, {Burke}, {Casey}, {Crawford},
  {Dencheva}, {Ely}, {Jenness}, {Labrie}, {Lim}, {Pierfederici}, {Pontzen},
  {Ptak}, {Refsdal}, {Servillat}, \& {Streicher}}]{2013A&A...558A..33A}
{Astropy Collaboration}, {Robitaille}, T.~P., {Tollerud}, E.~J., {et~al.} 2013,
  \aap, 558, A33

\bibitem[{{Bonnarel} {et~al.}(2000){Bonnarel}, {Fernique}, {Bienaym{\'e}},
  {Egret}, {Genova}, {Louys}, {Ochsenbein}, {Wenger}, \&
  {Bartlett}}]{2000A&AS..143...33B}
{Bonnarel}, F., {Fernique}, P., {Bienaym{\'e}}, O., {et~al.} 2000, \aaps, 143,
  33

\bibitem[{{Bundy} {et~al.}(2015){Bundy}, {Bershady}, {Law}, {Yan}, {Drory},
  {MacDonald}, {Wake}, {Cherinka}, {S{\'a}nchez-Gallego}, {Weijmans}, {Thomas},
  {Tremonti}, {Masters}, {Coccato}, {Diamond-Stanic}, {Arag{\'o}n-Salamanca},
  {Avila-Reese}, {Badenes}, {Falc{\'o}n-Barroso}, {Belfiore}, {Bizyaev},
  {Blanc}, {Bland-Hawthorn}, {Blanton}, {Brownstein}, {Byler}, {Cappellari},
  {Conroy}, {Dutton}, {Emsellem}, {Etherington}, {Frinchaboy}, {Fu}, {Gunn},
  {Harding}, {Johnston}, {Kauffmann}, {Kinemuchi}, {Klaene}, {Knapen},
  {Leauthaud}, {Li}, {Lin}, {Maiolino}, {Malanushenko}, {Malanushenko}, {Mao},
  {Maraston}, {McDermid}, {Merrifield}, {Nichol}, {Oravetz}, {Pan}, {Parejko},
  {Sanchez}, {Schlegel}, {Simmons}, {Steele}, {Steinmetz}, {Thanjavur},
  {Thompson}, {Tinker}, {van den Bosch}, {Westfall}, {Wilkinson}, {Wright},
  {Xiao}, \& {Zhang}}]{2015ApJ...798....7B}
{Bundy}, K., {Bershady}, M.~A., {Law}, D.~R., {et~al.} 2015, \apj, 798, 7

\bibitem[{{Chen} {et~al.}(2016){Chen}, {Shi}, {Tremonti}, {Bershady},
  {Merrifield}, {Emsellem}, {Jin}, {Huang}, {Fu}, {Wake}, {Bundy}, {Stark},
  {Lin}, {Argudo-Fernandez}, {Bergmann}, {Bizyaev}, {Brownstein}, {Bureau},
  {Chisholm}, {Drory}, {Guo}, {Hao}, {Hu}, {Li}, {Li}, {Lopes}, {Pan},
  {Riffel}, {Thomas}, {Wang}, {Westfall}, \& {Yan}}]{2016NatCo...713269C}
{Chen}, Y.-M., {Shi}, Y., {Tremonti}, C.~A., {et~al.} 2016, Nature
  Communications, 7, 13269

\bibitem[{{Comparato} {et~al.}(2007){Comparato}, {Becciani}, {Costa},
  {Larsson}, {Garilli}, {Gheller}, \& {Taylor}}]{2007PASP..119..898C}
{Comparato}, M., {Becciani}, U., {Costa}, A., {et~al.} 2007, \pasp, 119, 898

\bibitem[{{Gaia Collaboration} {et~al.}(2016){Gaia Collaboration}, {Prusti},
  {de Bruijne}, {Brown}, {Vallenari}, {Babusiaux}, {Bailer-Jones}, {Bastian},
  {Biermann}, {Evans}, \& et~al.}]{2016A&A...595A...1G}
{Gaia Collaboration}, {Prusti}, T., {de Bruijne}, J.~H.~J., {et~al.} 2016,
  \aap, 595, A1

\bibitem[{{Goodman} {et~al.}(2014){Goodman}, {Pepe}, {Blocker}, {Borgman},
  {Cranmer}, {Crosas}, {Di Stefano}, {Gil}, {Groth}, {Hedstrom}, {Hogg},
  {Kashyap}, {Mahabal}, {Siemiginowska}, \& {Slavkovic}}]{2014PLSCB..10E3542G}
{Goodman}, A., {Pepe}, A., {Blocker}, A.~W., {et~al.} 2014, PLoS Computational
  Biology, 10, e1003542

\bibitem[{{Goodman}(2012)}]{2012AN....333..505G}
{Goodman}, A.~A. 2012, Astronomische Nachrichten, 333, 505

\bibitem[{{Hassan} \& {Fluke}(2011)}]{2011PASA...28..150H}
{Hassan}, A. \& {Fluke}, C.~J. 2011, \pasa, 28, 150

\bibitem[{Hunter(2007)}]{Hunter:2007}
Hunter, J.~D. 2007, Computing In Science \& Engineering, 9, 90

\bibitem[{{Jin} {et~al.}(2016){Jin}, {Chen}, {Shi}, {Tremonti}, {Bershady},
  {Merrifield}, {Emsellem}, {Fu}, {Wake}, {Bundy}, {Lin}, {Argudo-Fernandez},
  {Huang}, {Stark}, {Storchi-Bergmann}, {Bizyaev}, {Brownstein}, {Chisholm},
  {Guo}, {Hao}, {Hu}, {Li}, {Li}, {Masters}, {Malanushenko}, {Pan}, {Riffel},
  {Roman-Lopes}, {Simmons}, {Thomas}, {Wang}, {Westfall}, \&
  {Yan}}]{2016MNRAS.463..913J}
{Jin}, Y., {Chen}, Y., {Shi}, Y., {et~al.} 2016, \mnras, 463, 913

\bibitem[{Jones {et~al.}(2001)Jones, Oliphant, Peterson, {et~al.}}]{citescipy}
Jones, E., Oliphant, T., Peterson, P., {et~al.} 2001, {SciPy}: Open source
  scientific tools for {Python}, [Online; accessed 2016-01-15]

\bibitem[{{Kent}(2013)}]{2013PASP..125..731K}
{Kent}, B.~R. 2013, \pasp, 125, 731

\bibitem[{{Naiman}(2016)}]{2016A&C....15...50N}
{Naiman}, J.~P. 2016, Astronomy and Computing, 15, 50

\bibitem[{P\'erez \& Granger(2007)}]{PER-GRA:2007}
P\'erez, F. \& Granger, B.~E. 2007, Computing in Science and Engineering, 9, 21

\bibitem[{{S{\'a}nchez} {et~al.}(2012){S{\'a}nchez}, {Kennicutt}, {Gil de Paz},
  {van de Ven}, {V{\'{\i}}lchez}, {Wisotzki}, {Walcher}, {Mast}, {Aguerri},
  {Albiol-P{\'e}rez}, {Alonso-Herrero}, {Alves}, {Bakos}, {Bart{\'a}kov{\'a}},
  {Bland-Hawthorn}, {Boselli}, {Bomans}, {Castillo-Morales}, {Cortijo-Ferrero},
  {de Lorenzo-C{\'a}ceres}, {Del Olmo}, {Dettmar}, {D{\'{\i}}az}, {Ellis},
  {Falc{\'o}n-Barroso}, {Flores}, {Gallazzi}, {Garc{\'{\i}}a-Lorenzo},
  {Gonz{\'a}lez Delgado}, {Gruel}, {Haines}, {Hao}, {Husemann},
  {Igl{\'e}sias-P{\'a}ramo}, {Jahnke}, {Johnson}, {Jungwiert}, {Kalinova},
  {Kehrig}, {Kupko}, {L{\'o}pez-S{\'a}nchez}, {Lyubenova}, {Marino},
  {M{\'a}rmol-Queralt{\'o}}, {M{\'a}rquez}, {Masegosa}, {Meidt},
  {Mendez-Abreu}, {Monreal-Ibero}, {Montijo}, {Mour{\~a}o}, {Palacios-Navarro},
  {Papaderos}, {Pasquali}, {Peletier}, {P{\'e}rez}, {P{\'e}rez}, {Quirrenbach},
  {Rela{\~n}o}, {Rosales-Ortega}, {Roth}, {Ruiz-Lara},
  {S{\'a}nchez-Bl{\'a}zquez}, {Sengupta}, {Singh}, {Stanishev}, {Trager},
  {Vazdekis}, {Viironen}, {Wild}, {Zibetti}, \&
  {Ziegler}}]{2012A&A...538A...8S}
{S{\'a}nchez}, S.~F., {Kennicutt}, R.~C., {Gil de Paz}, A., {et~al.} 2012,
  \aap, 538, A8

\bibitem[{{Sciacca} {et~al.}(2015){Sciacca}, {Becciani}, {Costa}, {Vitello},
  {Massimino}, {Bandieramonte}, {Krokos}, {Riggi}, {Pistagna}, \&
  {Taffoni}}]{2015A&C....11..146S}
{Sciacca}, E., {Becciani}, U., {Costa}, A., {et~al.} 2015, Astronomy and
  Computing, 11, 146

\bibitem[{{SDSS Collaboration} {et~al.}(2016){SDSS Collaboration}, {Albareti},
  {Allende Prieto}, {Almeida}, {Anders}, {Anderson}, {Andrews},
  {Aragon-Salamanca}, {Argudo-Fernandez}, {Armengaud}, \&
  et~al.}]{2016arXiv160802013S}
{SDSS Collaboration}, {Albareti}, F.~D., {Allende Prieto}, C., {et~al.} 2016,
  ArXiv e-prints [\eprint[arXiv]{1608.02013}]

\bibitem[{{Subba Rao} {et~al.}(2008){Subba Rao}, {Arag{\'o}n-Calvo}, {Chen},
  {Quashnock}, {Szalay}, \& {York}}]{2008NJPh...10l5015S}
{Subba Rao}, M.~U., {Arag{\'o}n-Calvo}, M.~A., {Chen}, H.~W., {et~al.} 2008,
  New Journal of Physics, 10, 125015

\bibitem[{{Taylor}(2005)}]{2005ASPC..347...29T}
{Taylor}, M.~B. 2005, in Astronomical Society of the Pacific Conference Series,
  Vol. 347, Astronomical Data Analysis Software and Systems XIV, ed.
  P.~{Shopbell}, M.~{Britton}, \& R.~{Ebert}, 29

\bibitem[{{Vanderplas} {et~al.}(2012){Vanderplas}, {Connolly}, {Ivezi{\'c}}, \&
  {Gray}}]{astroML}
{Vanderplas}, J., {Connolly}, A., {Ivezi{\'c}}, {\v Z}., \& {Gray}, A. 2012, in
  Conference on Intelligent Data Understanding (CIDU), 47 --54

\bibitem[{{Verley} {et~al.}(2007){Verley}, {Leon}, {Verdes-Montenegro},
  {Combes}, {Sabater}, {Sulentic}, {Bergond}, {Espada}, {Garc{\'{\i}}a},
  {Lisenfeld}, \& {Odewahn}}]{2007A&A...472..121V}
{Verley}, S., {Leon}, S., {Verdes-Montenegro}, L., {et~al.} 2007, \aap, 472,
  121

\bibitem[{Walt {et~al.}(2011)Walt, Colbert, \&
  Varoquaux}]{:/content/aip/journal/cise/13/2/10.1109/MCSE.2011.37}
Walt, S. v.~d., Colbert, S.~C., \& Varoquaux, G. 2011, Computing in Science \&
  Engineering, 13, 22

\bibitem[{{Yan} {et~al.}(2016){Yan}, {Bundy}, {Law}, {Bershady}, {Andrews},
  {Cherinka}, {Diamond-Stanic}, {Drory}, {MacDonald}, {S{\'a}nchez-Gallego},
  {Thomas}, {Wake}, {Weijmans}, {Westfall}, {Zhang}, {Arag{\'o}n-Salamanca},
  {Belfiore}, {Bizyaev}, {Blanc}, {Blanton}, {Brownstein}, {Cappellari},
  {D'Souza}, {Emsellem}, {Fu}, {Gaulme}, {Graham}, {Goddard}, {Gunn},
  {Harding}, {Jones}, {Kinemuchi}, {Li}, {Li}, {Maiolino}, {Mao}, {Maraston},
  {Masters}, {Merrifield}, {Oravetz}, {Pan}, {Parejko}, {Sanchez}, {Schlegel},
  {Simmons}, {Thanjavur}, {Tinker}, {Tremonti}, {van den Bosch}, \&
  {Zheng}}]{2016AJ....152..197Y}
{Yan}, R., {Bundy}, K., {Law}, D.~R., {et~al.} 2016, \aj, 152, 197

\end{thebibliography}

\end{document}